%% file: start.tex
\documentclass{format170x240multiauthor}
\usepackage[T1]{fontenc}
\usepackage{makeidx}
   \makeindex
\usepackage{amsmath}
\usepackage{textcomp}
\usepackage{graphicx}
\usepackage{cite}
\usepackage{url}
\begin{document}

\pagenumbering{Roman}
\raggedbottom

\pagenumbering{arabic}

\include{steps}


{\cleardoubleemptypage
  \addtocontents{toc}{%
    \protect\tocline
      {0}%
      {}%
      {\indexname}%
      {\sffamily\arabic{page}}}%
\footnotesize
\printindex}
\end{document}

%% file: steps.tex
\chapter{Nonlinear dynamics of surface steps}
\label{ch1}
\chapterauthor[Joachim Krug]
{Joachim Krug}

\section{Introduction}
\label{Krug:Intro}

Surface steps are key elements in the dynamics of a crystal surface
below its thermodynamic roughening transition, because
they constitute long-lived structural defects which are nevertheless highly
mobile and prone to strong fluctuations \cite{Krug:Williams2004}. 
The description of surface
morphology evolution in terms of the thermodynamics and kinetics of 
steps goes back at least half a century \cite{Krug:Burton1951}. During
the past few decades the subject has experienced a significant revival 
due to the availability of imaging methods such as scanning
tunneling microscopy, which allow for a direct visualization of  
step conformation and step motion on the nanoscale; 
see \cite{Krug:Jeong99,Krug:Giesen01,Krug:Michely2004,Krug:PierreLouis2005,Krug:Evans2006} 
for recent reviews. 
In this chapter I will focus specifically on cases
where steps have been found to display complex \textit{dynamic} behavior,
such as oscillatory shape evolution under constant driving. 

The examples to be discussed below can be naturally organized
according to the underlying topology of the step configurations: We 
first consider driven single-layer islands (closed step loops), 
and then vicinal surfaces (arrays of parallel steps). A certain
familiarity with the basic thermodynamics and kinetics of crystal surfaces
is assumed; for an elementary introduction the reader may consult
\cite{Krug:Krug2005}.  

\section{Electromigration-driven islands and voids}
\label{Krug:Islands}

\begin{figure}[t]
\def\capfrac{1}
\centerline{
\includegraphics[width=.7\textwidth]{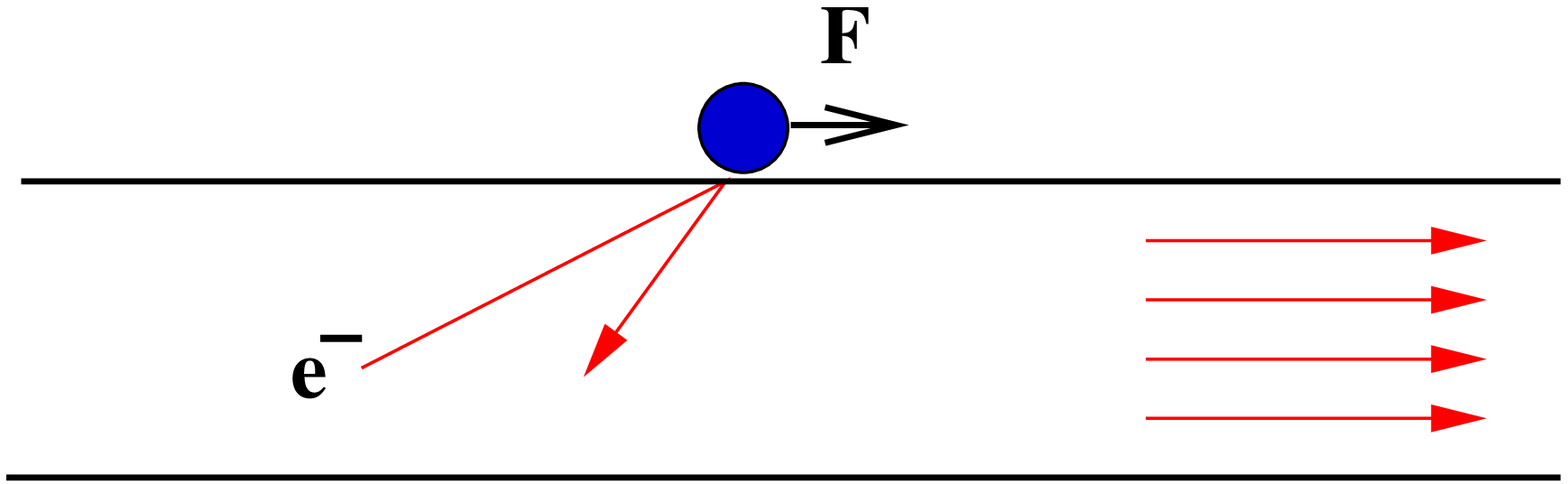}}
\Caption{Schematic of the microscopic origin of the electromigration force:
Conduction electrons scattering off an adatom give rise to a transfer of momentum
in the direction of the current flow.}\label{fig:emig}
\end{figure}

Electromigration is the directed transport of matter in a current-carrying
material, which is caused (primarily) by the scattering of conduction electrons
off defects such as interstitials or atoms adsorbed on the surface
(henceforth referred to as \textit{adatoms}), see Fig.\ref{fig:emig}
for a sketch. Much of the work on electromigration
has been motivated by its importance as a damage mechanism limiting the lifetime
of integrated circuits \cite{Krug:Tu03}. Because electromigration forces are small
compared to the typical energy barriers involved in the thermal diffusion of atoms,
the direct observation of electromigration effects in real time on atomistic length
scales is difficult (see, however, \cite{Krug:Bondarchuk2007} for recent progress in this direction). 
In this chapter the electromigration force will be used as a conceptually simple way of driving a system of surface
steps out of equilibrium, giving rise to surprisingly complex dynamical behavior.

\subsection{Electromigration of single layer islands}

Two-dimensional single-layer islands are the simplest nanoscale structures that appear on a surface
during the early stages of thin film growth, when the amount of deposited material is a small fraction
of a monolayer \cite{Krug:Michely2004}. Because of their small size, such islands display considerable
shape fluctuations already in thermal equilibrium, which may cause diffusive motion of the island
as a whole \cite{Krug:Giesen01}. The electromigration-induced drift of
single-layer islands on the Si(111) surface was observed experimentally by M\'etois and collaborators 
in 1999 \cite{Krug:Metois1999}. In the following we summarize recent theoretical work on 
this problem, which is based on a continuum formulation due to Pierre-Louis and Einstein 
\cite{Krug:PierreLouis2000}.

We focus here on the simplest case where the motion of atoms is restricted to 
the boundary of the islands, such that the island area is conserved\footnote{A nonconserved
situation where the step exchanges atoms with the terrace is treated below in 
Sect.\ref{Nonlocal}.}. 
Then the local normal velocity $v_n$ of the island boundary satisfies a continuity
equation,
\begin{equation}
\label{vn}
v_n = - \frac{\partial}{\partial s} j = \frac{\partial}{\partial s} \sigma \left[ \frac{\partial}{\partial s}
(\tilde \gamma \kappa) - F_t \right],
\end{equation}
where $s$ denotes the arclength measured along the island contour. The mass current $j$ along the island
boundary is proportional to the step edge mobility $\sigma$, and it is driven by capillary forces
and the tangential (to the boundary) component $F_t$ of the electromigration force. 
The capillary force, in turn, is given by the tangential gradient of the edge chemical potential,
which is the product of the edge stiffness $\tilde \gamma$ and the edge curvature $\kappa$. The stiffness
$\tilde \gamma$ is derived from the edge free energy per unit length $\gamma$ according to 
$\tilde \gamma = \gamma + \gamma''$, where primes denote derivatives with respect to the orientation
angle of the edge. In the absence of external forces ($F_t = 0$) Eq.(\ref{vn}) guarantees the relaxation
of the island to its equilibrium shape characterized by $\tilde \gamma \kappa = \mathrm{const.}$
\cite{Krug:Michely2004}. Throughout this section the electromigration force is assumed to be 
constant in magnitude and direction. This implies that 
\begin{equation}
\label{Ft}
F_t = F_0 \cos \theta, 
\end{equation}
where $\theta$ denotes the angle between the boundary and the direction of the force.

\begin{figure}[t]
\def\capfrac{1}
\includegraphics[width=.6\textwidth,angle=-90]{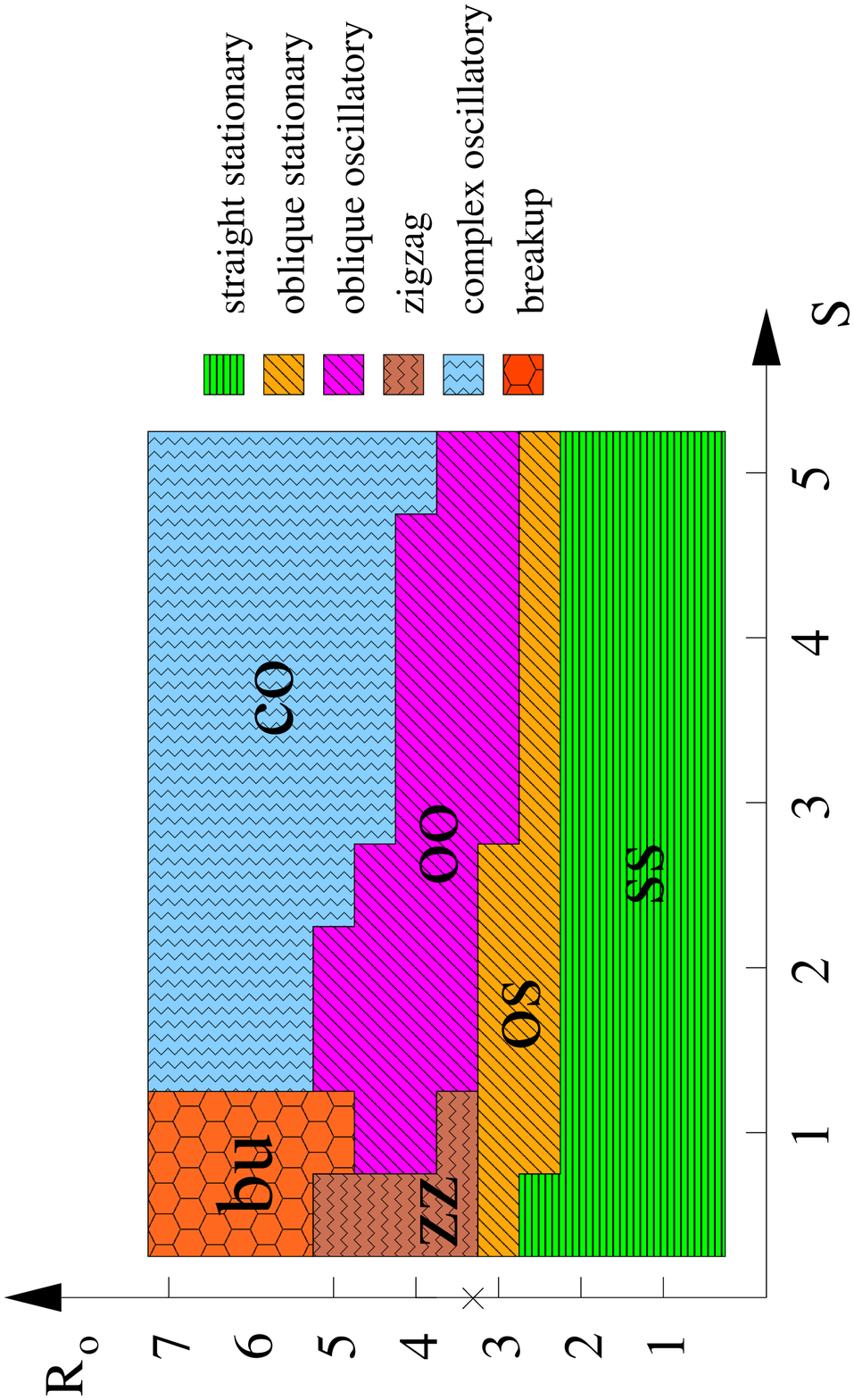}
\Caption{Numerically generated phase diagram of island migration modes as a function
of the anisotropy strength $S$, as defined in (\ref{sigma}), and the dimensionless
island radius $R_0 = R/l_E$. In the regions denoted by \textbf{zz} and \textbf{oo}
the island shape oscillates periodically, while in the \textbf{co} region the 
behavior is irregular, possibly chaotic. The cross on the $R_0$-axis indicates
the bifurcation from circular to elongated shapes in the isotropic case at
the critical radius (\ref{Rc}). The phase diagram is based on a grid
of resolution $0.5 \times 0.5$ in the $S$-$R_0$-plane.}\label{fig:Phasediag}
\end{figure}

In the absence of crystalline anisotropy the material parameters $\sigma$ and $\tilde \gamma$ in
(\ref{vn}) are constants, and it is straightforward to check that Eqs.(\ref{vn},\ref{Ft}) are solved
by a circle of arbitrary radius $R$ moving at constant speed $V = \sigma F_0/R$ \cite{Krug:Ho1970}.
Linear stability analysis of the circular solution shows that it becomes unstable at a critical
radius \cite{Krug:Suo1996}
\begin{equation}
\label{Rc}
R_c \approx 3.26 \; l_E,
\end{equation}
where the characteristic length scale obtained by non-dimensionalizing (\ref{vn}) reads
\begin{equation}
\label{lE}
l_E = \sqrt{\tilde \gamma/F_0}.
\end{equation}
Beyond the linear instability of the circular solution one finds a family of stationary
shapes which are elongated in the direction of the force and become increasingly
sensitive to breakup with increasing size \cite{Krug:Kuhn2005a,Krug:Kuhn2007}.  

\begin{figure}[t]
\def\capfrac{1}
\centerline{
\includegraphics[width=.7\textwidth]{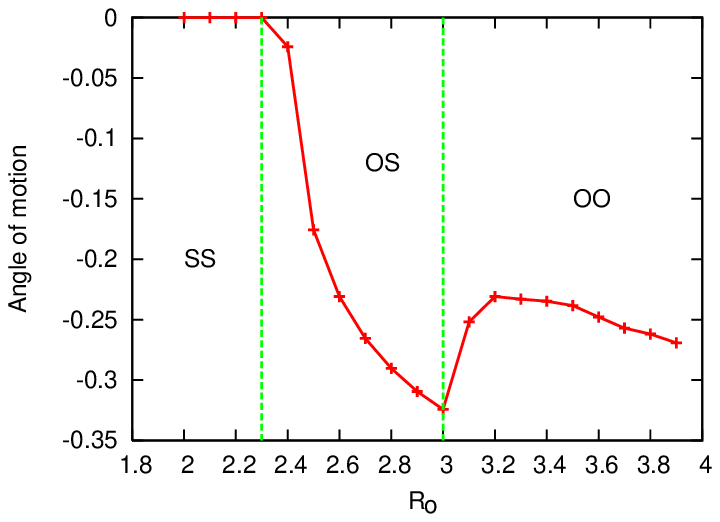}}
\Caption{Angle enclosed by the direction of island motion and the direction of the applied force
as a function of the scaled island radius for $S=2$. The transitions between the different
phases in Fig.\ref{fig:Phasediag} are manifest as slope discontinuities in this graph.}
\label{fig:Angle}
\end{figure}

The effect of crystalline anisotropy in the mobility $\sigma$ was explored,
mostly numerically, in 
\cite{Krug:Kuhn2005a,Krug:Kuhn2005b}. Using the expression \cite{Krug:Schimschak2000}
\begin{equation}
\label{sigma}
\sigma(\theta) = \sigma_0 [1 + S \cos^2 (n \theta)],
\end{equation}
where $2n$ denotes the number of symmetry axes, a surprisingly rich
phase diagram of migration modes was obtained in the plane spanned by
the anisotropy strength $S$ and the dimensionless island radius 
$R_0 = R/l_E$ for the case of sixfold anisotropy ($n=3$), see Fig.\ref{fig:Phasediag}.  
In these calculations the force was oriented along a direction of maximal
mobility. 

\begin{figure}[t]
\def\capfrac{1}
\includegraphics[width=0.9\textwidth]{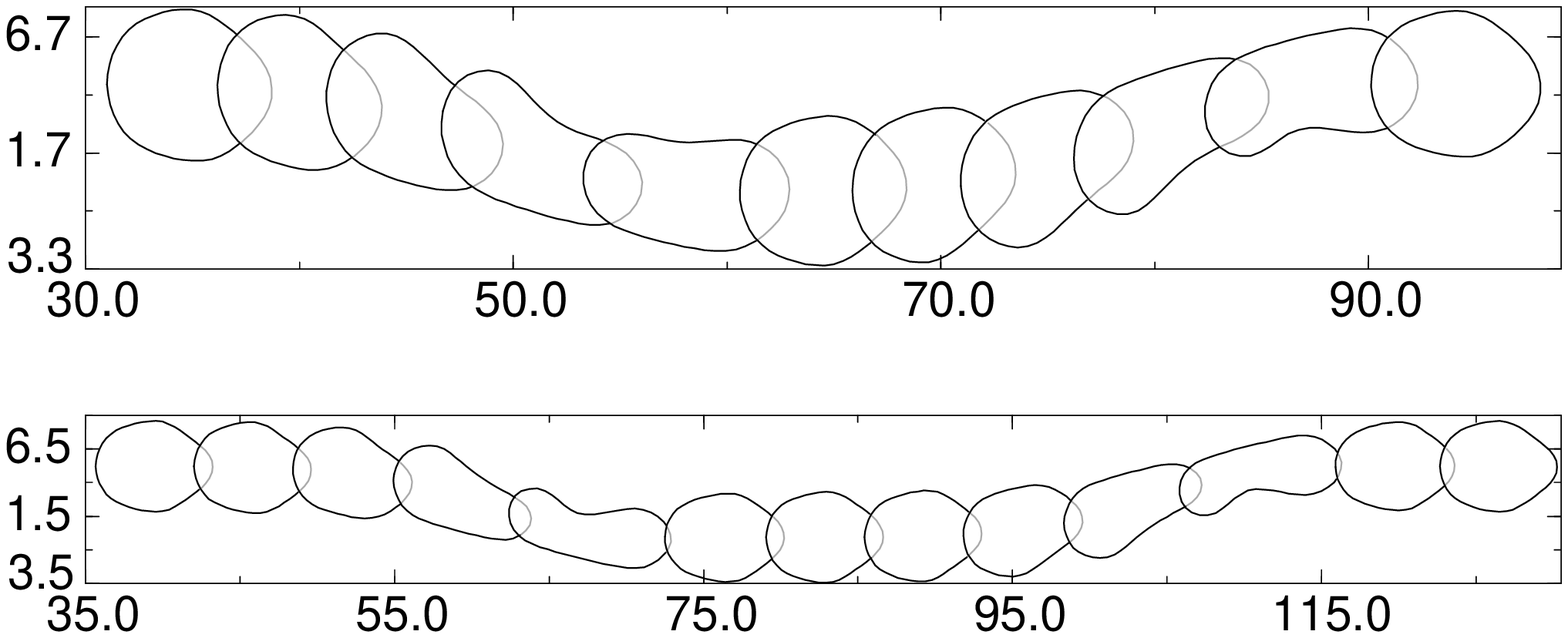}
\Caption{Oscillatory island motion in the zig-zag phase of the phase
diagram. Parameters are $R_0 = 3.5, S=0.5$ for the upper panel and
$R_0 = 3.5, S=1$ for the lower panel. All lengths are measured in units
of $l_E$. }\label{fig:zigzag}
\end{figure}

For small $R_0$ the dynamics is dominated by capillarity and the island shape is 
close to the equilibrium shape. The island moves at constant speed in the direction 
of the applied force (\textbf{ss} = straight stationary motion). With increasing
size a bifurcation to a regime of oblique stationary (\textbf{os}) motion occurs, in which 
the symmetry with respect to the force direction is spontaneously broken. A suitable
order parameter for this bifurcation is the angle between the direction of force and
the direction of motion (Fig.\ref{fig:Angle}). Increasing the radius further another
bifurcation occurs to a phase in which the obliquely moving island displays periodic
shape oscillations (the \textbf{oo} phase). At smaller values of $S$ the island performs
an oscillatory zig-zag motion which is directed along the applied force on average
(Fig.\ref{fig:zigzag}). 

\begin{figure}[t]
\def\capfrac{1}
\centerline{
\includegraphics[width=.8\textwidth]{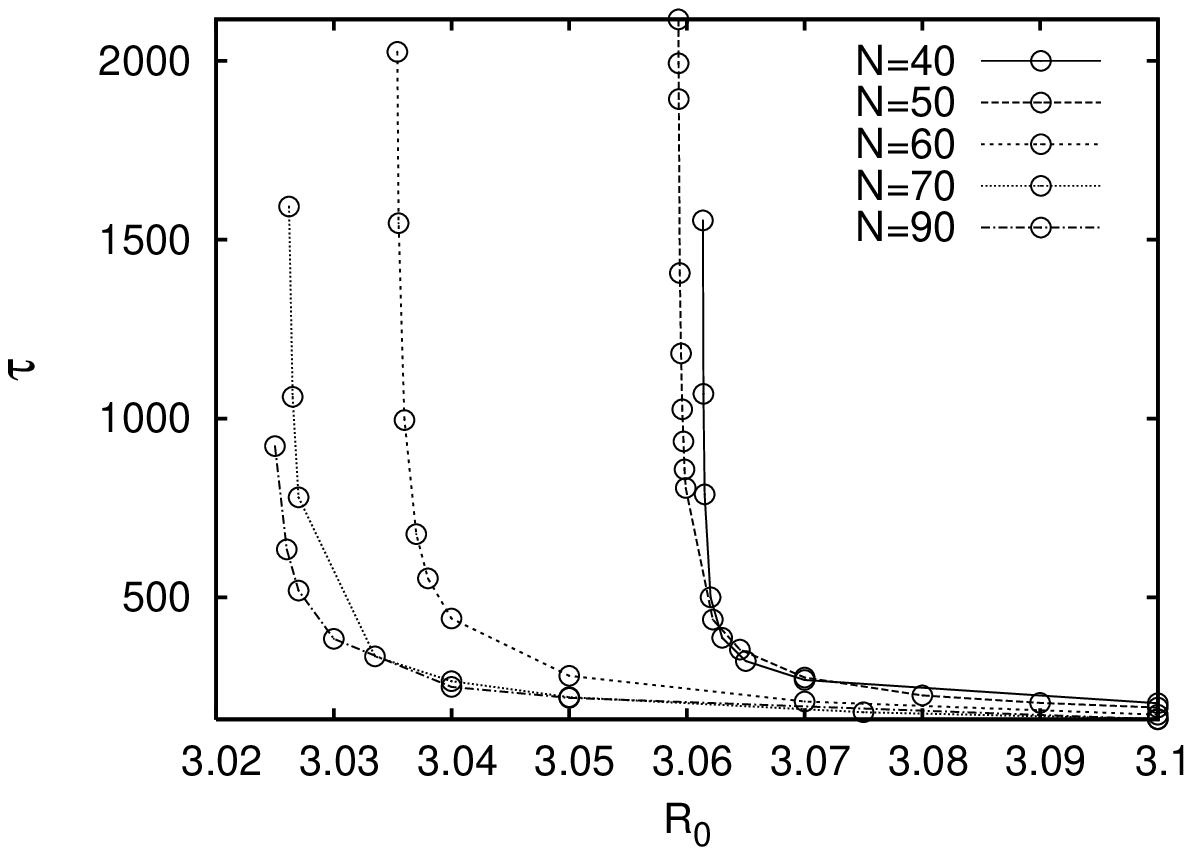}}
\Caption{Period $\tau$ of the shape oscillation near the transition from the 
\textbf{oo} to the \textbf{os} phase. Different curves show results obtained
for different numbers $N$ of discretization points in the numerical solution,
with $N$ increasing from right to left
(from \cite{Krug:Kuhn2007}).}
\label{fig:Period}
\end{figure}

A clear signature of the transition from stationary oblique to oscillatory behavior
shows up in the angle of island migration (Fig.\ref{fig:Angle}). In addition,
we observe that the period $\tau$ of the shape oscillation diverges as the critical
radius $R_0^{\textbf{oo}}$ of the transition is approached from above (Fig.\ref{fig:Period}). Although
the data show some dependence on the number of discretization points, a power law
fit indicates that the period diverges as
\begin{equation}
\label{T}
\tau \sim (R_0 - R_0^{\textbf{oo}})^{-2.5}.
\end{equation}

\begin{figure}[t]
\def\capfrac{1}
\centerline{
\includegraphics[width=.8\textwidth]{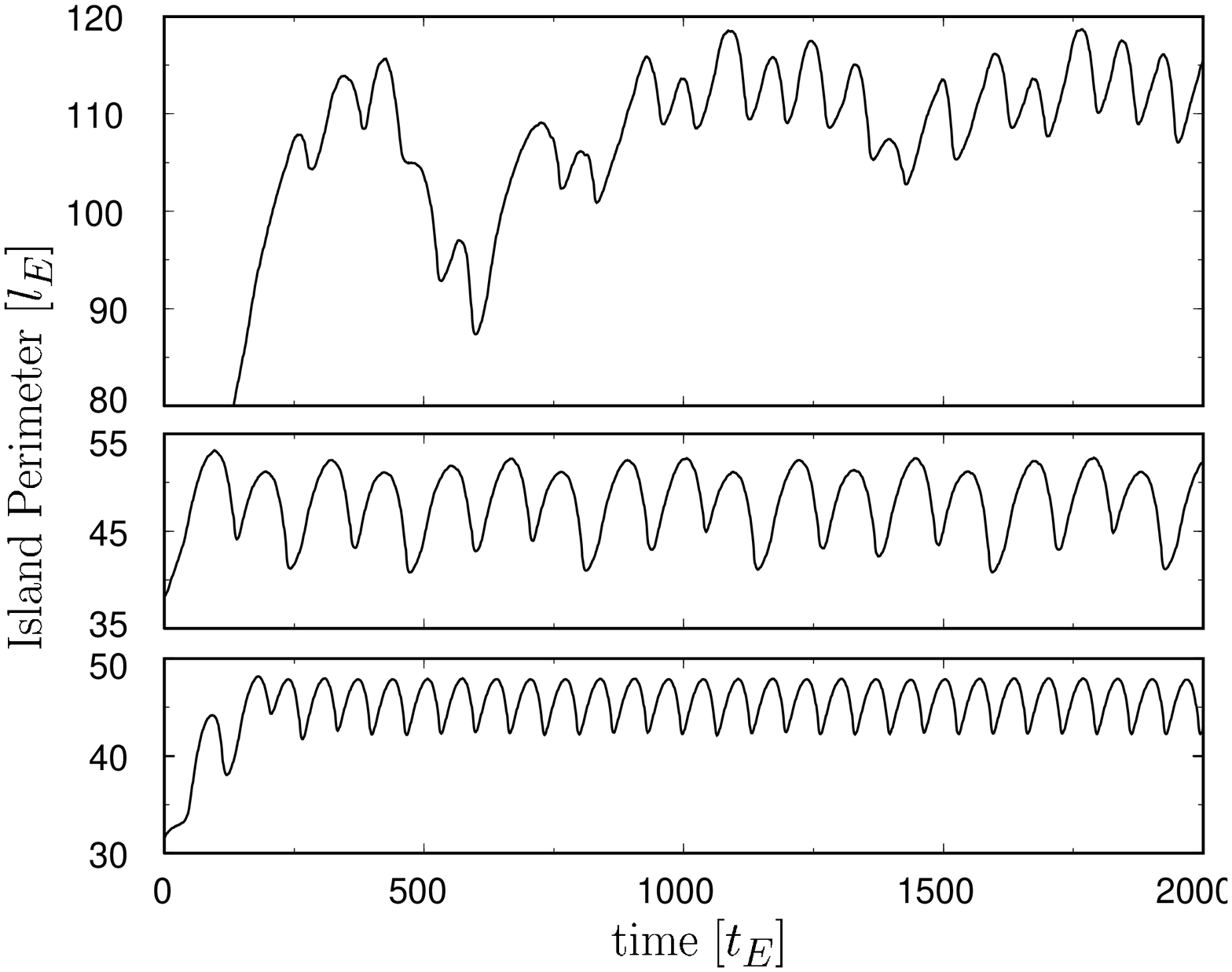}}
\Caption{Time series of the island perimeter, measured in units of $l_E$.
From bottom to top, parameters are $S=2, R_0 = 5$; $S=5, R_0 = 5$; and
$S=5, R_0 = 6.5$. Time is measured in units of $t_E = l_E^4/(\sigma_0 \tilde \gamma)$.}
\label{fig:time}
\end{figure}

\begin{figure}[t]
\def\capfrac{1}
\centerline{
\includegraphics[width=.75\textwidth]{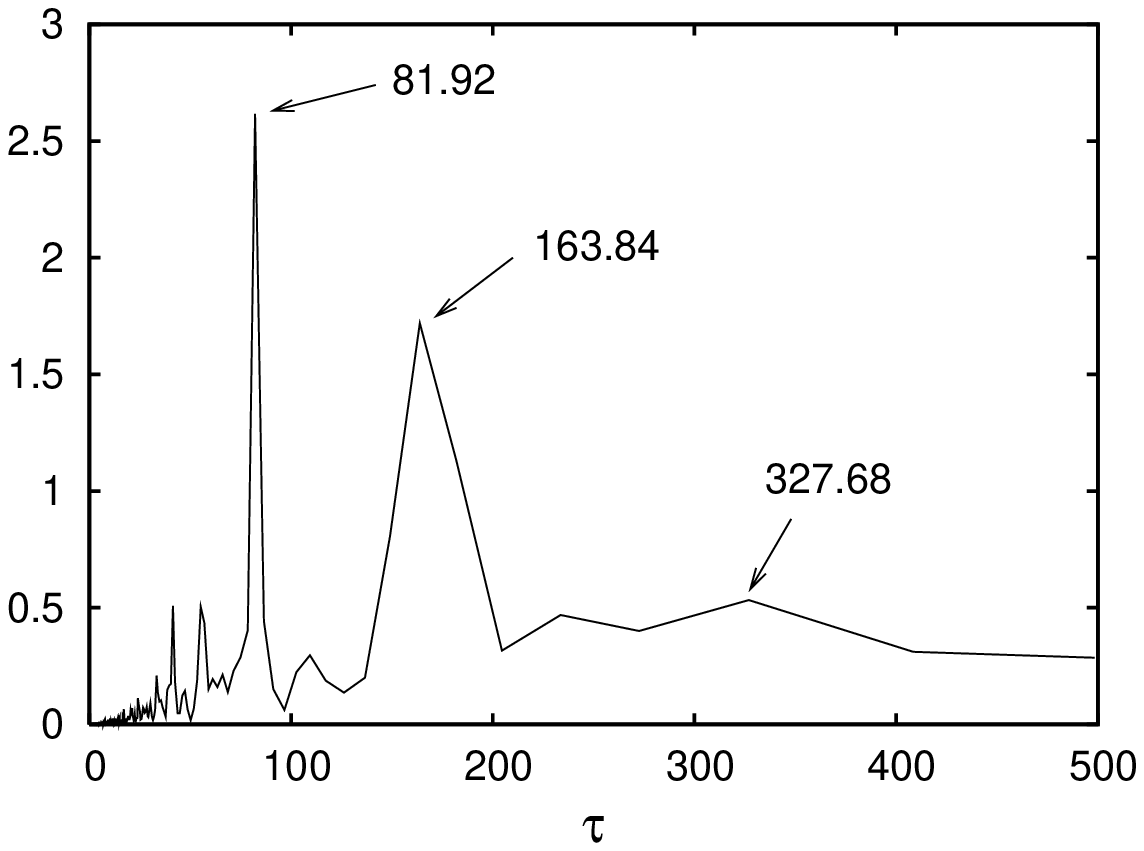}}
\Caption{Fourier spectrum of the island perimeter time series for
$S=3$ and $R_0 = 6$, plotted as a function of the
period $\tau = 2 \pi/\omega$  
(from \cite{Krug:Kuhn2007}).}\label{fig:Fourier}
\end{figure}

Increasing the island size further the oscillations become increasingly
irregular. This is illustrated in Fig.\ref{fig:time} by the time series
of the island perimeter. The uppermost curve in the figure displays large
scale fluctuations which can be traced back to reversals of the direction
of island motion which occur at irregular intervals \cite{Krug:Kuhn2005b}.
The Fourier spectrum of such a time series is broad and shows clear signatures
of period doubling (Fig.\ref{fig:Fourier}). 

\subsection{Continuum vs. discrete modeling}

We have seen in the preceding section that electromigration-driven islands 
display a number of features which are consistent with the behavior of a
low-dimensional, non-linear dynamical system. This is remarkable, since
physically such an island consists of a large number of atoms which move
stochastically under the influence of thermal fluctuations and a very small
systematic force. 

\begin{figure}[t]
\def\capfrac{1}
\centerline{
\includegraphics[width=.8\textwidth]{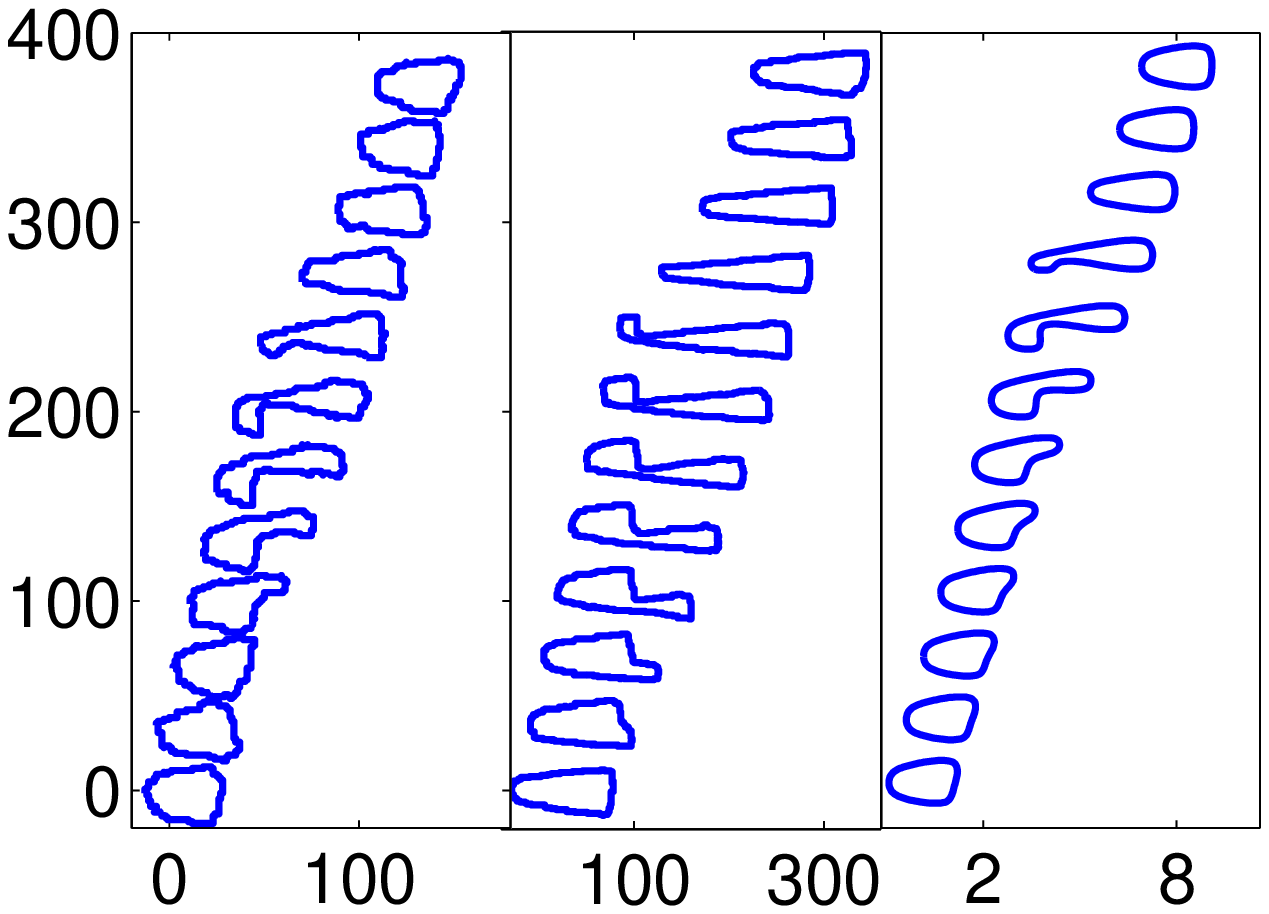}}
\Caption{Comparison of island shape evolution obtained from 
KMC simulations (left and middle columns) and numerical solution of
the continuum model (right column). The simulated islands consist
of 1000 atoms in the right column and 4000 atoms in the middle 
column. The left and right columns correspond to a temperature of
$T = 700$ K, while in the middle column $T=500$ K. In the left and middle columns
lengths are measured in units of the lattice constant, in the right column
in units of $l_E$.}\label{fig:KMC}
\end{figure}

In order to determine whether the phenomena predicted
on the basis of the deterministic continuum model (\ref{vn}) persist also
under experimentally realistic conditions, extensive Kinetic Monte Carlo
(KMC)
simulations were carried out using a lattice model that has been shown to
provide an accurate representation of metal surfaces\footnote{See 
\cite{Krug:Michely2004,Krug:Evans2006} for an overview of similar
models, and \cite{Krug:PierreLouis2000,Krug:Mehl2000} 
for earlier KMC-simulations of island electromigration.} 
such as Cu(100) \cite{Krug:Rusanen2006}. 
In a suitably chosen range of parameters, a regime of oscillatory motion
could be identified which shows dynamic behavior in good, essentially
quantitative agreement with the continuum model (Fig.\ref{fig:KMC}).

\begin{figure}[t]
\def\capfrac{1}
\centerline{
\includegraphics[width=.9\textwidth]{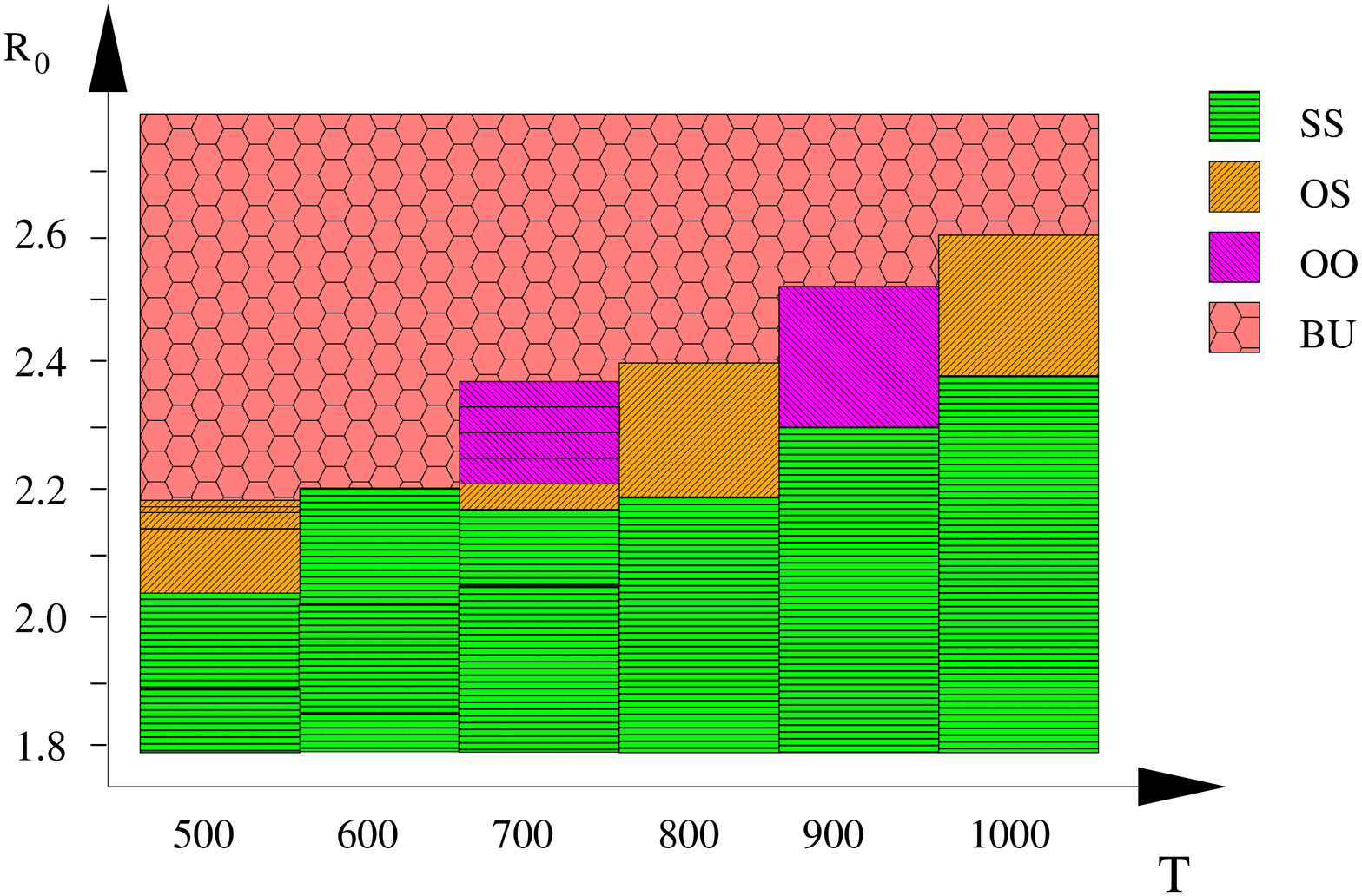}}
\Caption{Phase diagram of island migration modes obtained by
numerical solution of the continuum equations for a mobility and
stiffness of fourfold crystalline anisotropy. The temperature is
measured in Kelvin, and the anisotropy
increases with decreasing temperature. Temperature was varied
in steps of 100 K. Each rectangle corresponds to a single
value of $R_0$ and $T$, which is located in the center of the
rectangle. The cases $T = 500$K and $T = 700$K, which correspond to
the KMC simulations, were explored with higher resolution. The abbreviations
used for the different phases are explained in 
Fig.\ref{fig:Phasediag} (from \cite{Krug:Kuhn2007}).}
\label{fig:Phasediag2}
\end{figure}

For the comparison to KMC simulations, realistic expressions for the 
step edge mobility $\sigma$ and the stiffness $\tilde \gamma$ 
in (\ref{vn}) were derived and implemented. Both of these quantities
display a fourfold anisotropy on the fcc(100) surface. A rough exploration
of the full phase diagram, conducted within the continuum model, is 
depicted in Fig.\ref{fig:Phasediag2}. Since the physical parameter controlling
the anisotropy is the temperature $T$, with lower temperatures corresponding
to more pronounced anisotropy, the temperature axis in 
Fig.\ref{fig:Phasediag2} replaces the anisotropy axis in 
Fig.\ref{fig:Phasediag}. The regions displaying oscillatory behavior
without leading to island breakup are much more limited than in the case
of sixfold anisotropy. In particular, at $T = 500$ K no oscillatory regime
was found in the continuum model, despite the fact that oscillations are
seen in the KMC simulations at this temperature (middle column in 
Fig.\ref{fig:KMC}). This is one of the indications of a breakdown of the
continuum description at low temperatures which were reported in
\cite{Krug:Rusanen2006}.

\subsection{Nonlocal shape evolution: Two-dimensional voids}

Formally, the island electromigration problem described in the preceding sections
is largely equivalent to the problem of electromigration of cylindrical voids 
in a thin metallic
film. The formation, migration and shape evolution of such voids plays an important
part in the failure of metallic interconnects in integrated electronic circuits
\cite{Krug:Tu03}. In this context the size scale of interest is usually in the range
of micrometers, rather than nanometers, but on the level of the continuum description on which
(\ref{vn}) is based this difference is immaterial. 

\begin{figure}[t]
\def\capfrac{1}
\includegraphics[width=\textwidth]{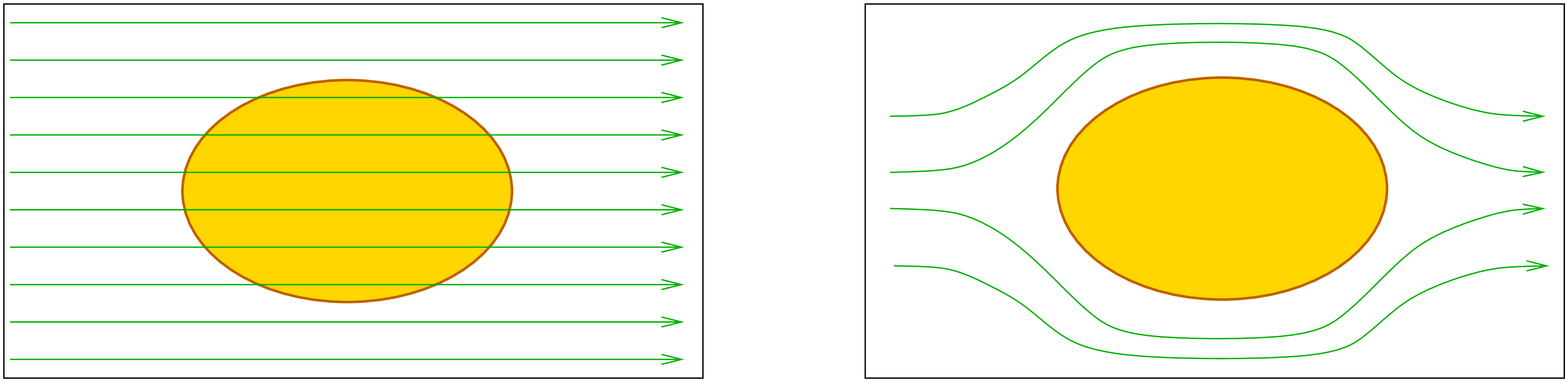}
\Caption{Comparison between the electromigration problem for islands (left) and voids (right).
Arrows indicate the flow of the electric current. The shape evolution problem on the left
is \textit{local}, whereas on the right one has to solve a \textit{nonlocal} moving boundary
value problem.}
\label{fig:2insel}
\end{figure}

A more relevant distinction is illustrated
in Fig.\ref{fig:2insel}: In the case of an island on top of a thick metallic substrate,
the disturbance of the electric current distribution in the bulk due to the presence of
the island can be neglected, and correspondingly the force $F_t$ in (\ref{vn}) can be
approximated by the simple constant expression (\ref{Ft}). On the other hand, in the
presence of an insulating void in a current-carrying film, the current is obviously forced
to flow around the void. As a consequence, the current distribution, and hence the distribution
of electromigration forces, is strongly dependent on the void shape itself, and the shape
evolution becomes a non-local moving boundary value problem for the 
electric potential \cite{Krug:Schimschak2000}.
It is possible to interpolate between the two cases depicted in Fig.\ref{fig:2insel} by
considering a conducting void and varying the conductivity ratio between the interior and
the exterior regions \cite{Krug:Suo1996}. 

Oscillatory shape evolution of two-dimensional voids was first observed numerically by Gungor
and Maroudas \cite{Krug:Gungor2000}. They considered edge voids located at the boundary
of a two-dimensional conducting strip. In the presence of crystalline anisotropy in the 
mobility of adatoms along the inner void surface, a transition from stationary to oscillatory
behavior occurs with increasing electromigration force or void area. Subsequent detailed
analysis has shown that this transition has the character of a Hopf bifurcation
\cite{Krug:Cho2008}. The experimental signature of oscillatory void evolution are rapid oscillations
in the resistance of the conductor, which have indeed been reported in the literature
\cite{Krug:Cho2005}.

\subsection{Nonlocal shape evolution: Vacancy islands with terrace diffusion}
\label{Nonlocal}

The exchange of atoms between the step and the surrounding terraces is
another source of nonlocality in the motion of the steps,
since it necessitates the solution of a moving
boundary value problem for the concentration of adatoms on the terraces
\cite{Krug:Burton1951,Krug:Krug2005}. A particular case in this class
of problems is the \textit{interior model} for the electromigration
of vacancy islands introduced in \cite{Krug:PierreLouis2000}, and
studied in detail in \cite{Krug:Hausser2007,Krug:Kuhn2007}.

\begin{figure}[t]
\def\capfrac{1}
\centerline{
\includegraphics[width=0.5\textwidth]{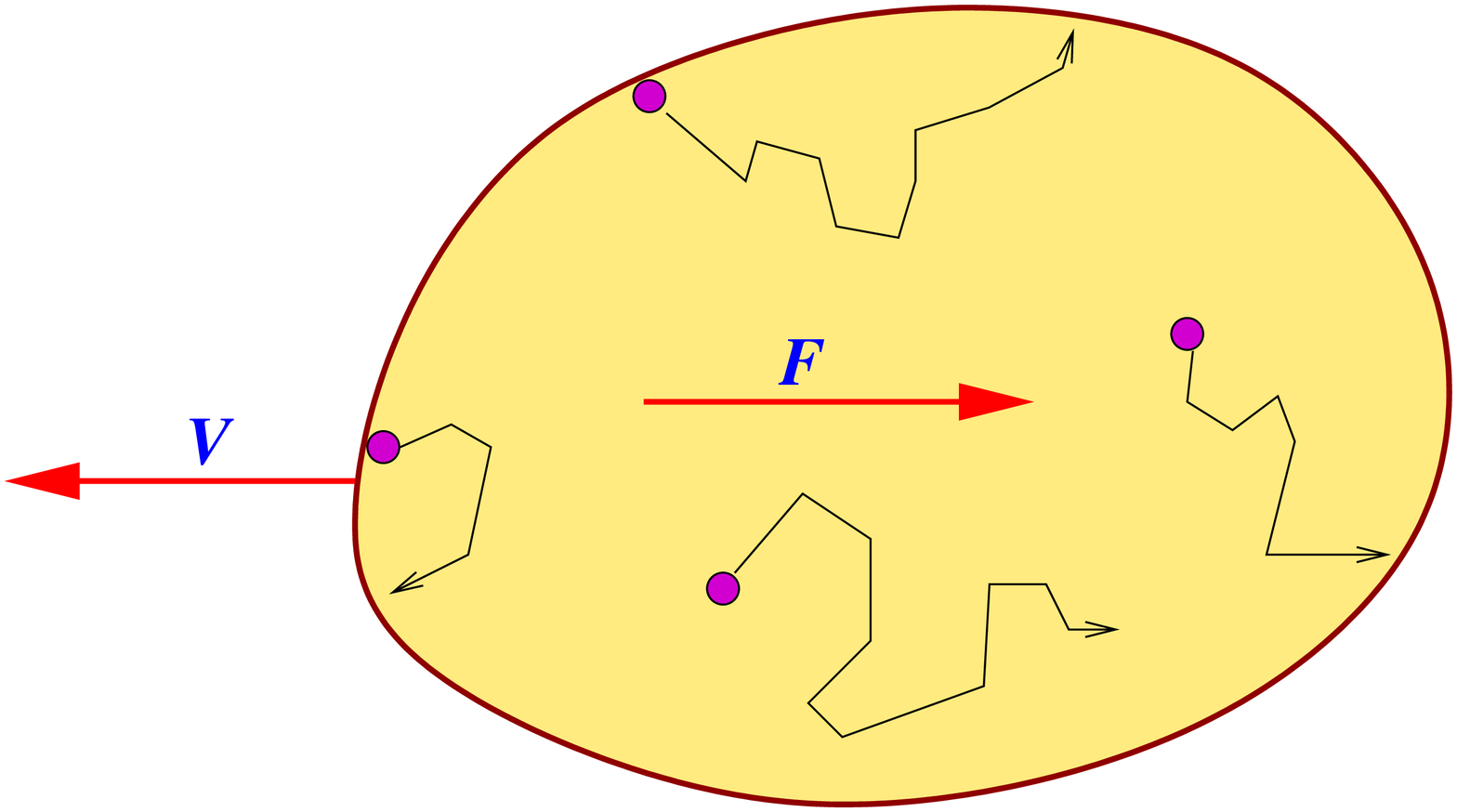}}
\Caption{Sketch of a vacancy island migrating by internal terrace diffusion.
The drift force leads to a net transport of material from the left to the
right, which implies island migration in the opposite direction.}
\label{fig:interior}
\end{figure}

As illustrated in Fig.\ref{fig:interior}, one considers a vacancy
island (i.e. a surface region which is one atomic height lower than
the surrounding terrace) bounded by an ascending step. Atoms can detach
from the step and diffuse across the island, but an energy barrier 
prevents atoms from entering the island from the exterior terrace.
This leads to a moving boundary value problem in the bounded
interior domain, where the adatom concentration $\rho(\vec r,t)$
satisfies the drift-diffusion equation 
\begin{equation}
\label{interioreq}
\frac{\partial \rho}{\partial t} = D \nabla^2 \rho - \frac{D}{k_\mathrm{B} T}
{\vec F} \cdot \nabla \rho
\end{equation}
with appropriate boundary conditions at the step edge (see 
\cite{Krug:Krug2005} for a general discussion). If the exchange of atoms with
the step edge is rapid, so that thermal equilibrium is maintained
at the boundary at all times, a circular stationary solution drifting
at constant speed against the force direction can be found 
\cite{Krug:PierreLouis2000}. 

From the perspective of nonlinear dynamics, an intriguing feature of this
problem is that the circular solution is linearly stable, although numerical
simulation of the fully nonlinear evolution shows that the circle
develops an instability
(leading eventually to the pinching off of a small island)
under finite perturbations \cite{Krug:Hausser2007}. The critical 
perturbation strength needed to trigger the instability 
decreases as the dimensionless island size, defined in this case by 
\begin{equation}
\label{Rcinterior}
R_0 = \frac{R}{\xi}, \;\;\; \xi = \frac{k_\mathrm{B} T}{\vert \vec F \vert},
\end{equation}
is increased by increasing either the force or the island size. A similar scenario
combining linear stability with nonlinear instability was previously
found in the problem of two-dimensional void migration 
\cite{Krug:Schimschak2000,Krug:Schimschak1998} as well as in the
dynamics of ionization fronts \cite{Krug:Meulenbroek2005,Krug:Ebert2007}.

The effects of crystalline anisotropy in this problem have not been explored
so far. However, in view of the results described in the preceding subsections,
it seems likely that oscillatory and other modes 
of complex shape evolution may arise in this case as well.

\begin{figure}[t]
\def\capfrac{1}
\centerline{
\includegraphics[width=0.8\textwidth]{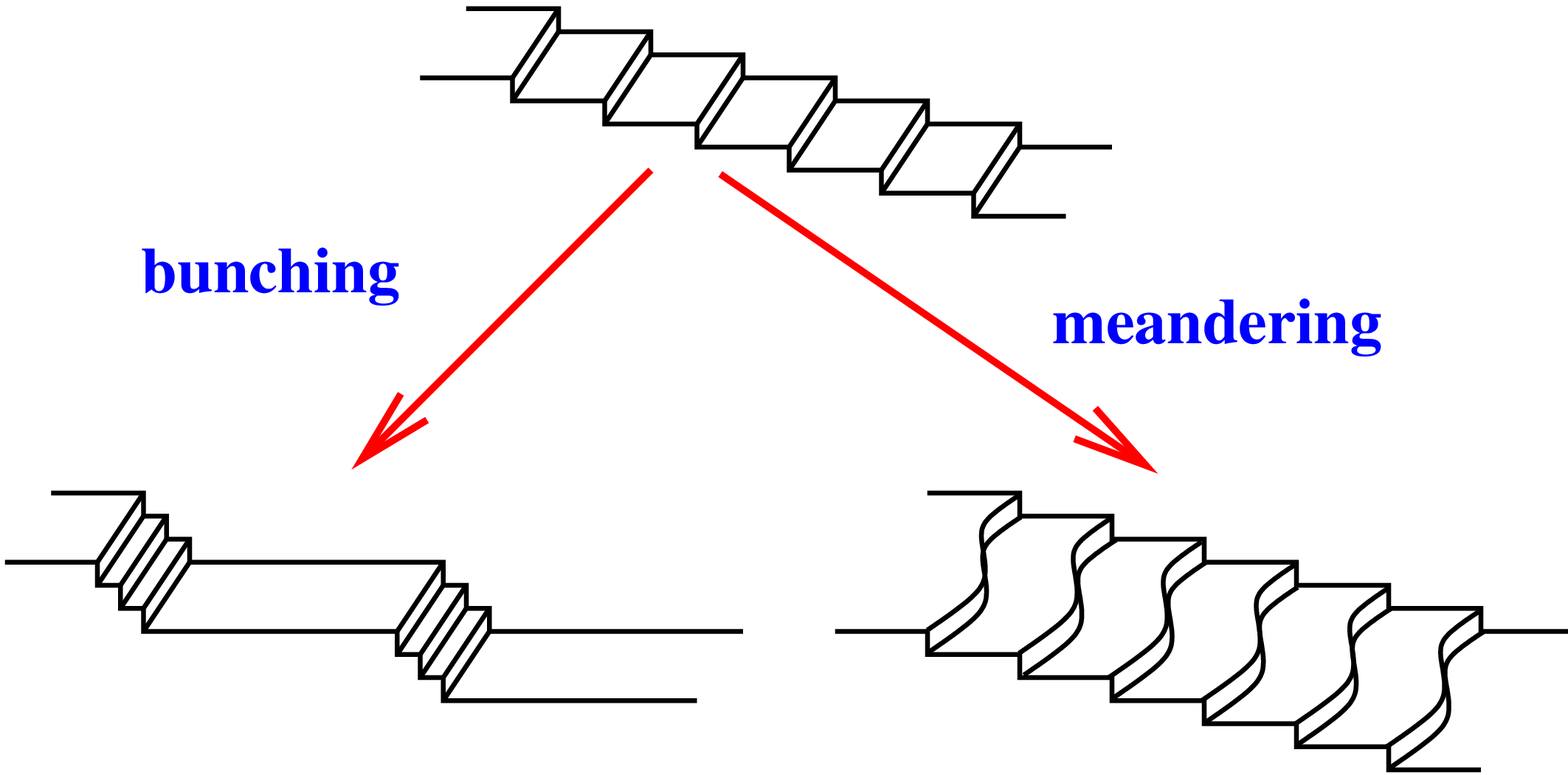}}
\Caption{Schematic of the two main morphological instabilities of a vicinal surface.}
\label{fig:vicinals}
\end{figure}

\section{Step bunching on vicinal surfaces}
\label{Krug:Bunches}

A vicinal surface is obtained by cutting a crystal at a small angle
relative to a high symmetry orientation, such that a staircase of
well-separated, atomic height steps forms. When such an array of steps
is set into motion by growing or sublimating the crystal, or by applying
an electromigration force on the adatoms, a variety of patterns emerges.

Quite generally, the pattern formation process can be understood as a competition
between the destabilizing effects of the external forces, and thermodynamic
forces arising from the step free energy and repulsive
step-step interactions, which act to restore the
equilibrium state of straight, equidistant steps. The resulting
instability scenarios have been studied extensively on the level
of linear stability analysis, see e.g. \cite{Krug:PierreLouis2003}. The two
basic modes of instability are illustrated in Fig.\ref{fig:vicinals}: In
\textit{step bunching} the individual steps remain straight but
the initially homogeneous step train breaks up into regions of high
step density (bunches) separated by wide terraces. By contrast,
in \textit{step meandering} the individual steps become wavy; often the
repulsive interactions between the steps then force the different steps
to meander in phase, such that an overall periodic surface 
corrugation perpendicular to the direction of vicinality results.
In some cases step bunching and step meandering have been observed
to coexist \cite{Krug:Neel2003,Krug:Yu2006}.

In the following some recent results on the \textit{nonlinear} evolution of 
step bunches will be summarized, focusing again on instances of complex
\textit{temporal} behavior of the step configurations.
For a discussion of the nonlinear dynamics of meandering steps we refer
to \cite{Krug:Krug2005a}. 

\begin{figure}[t]
\def\capfrac{1}
\centerline{
\includegraphics[width=0.5\textwidth]{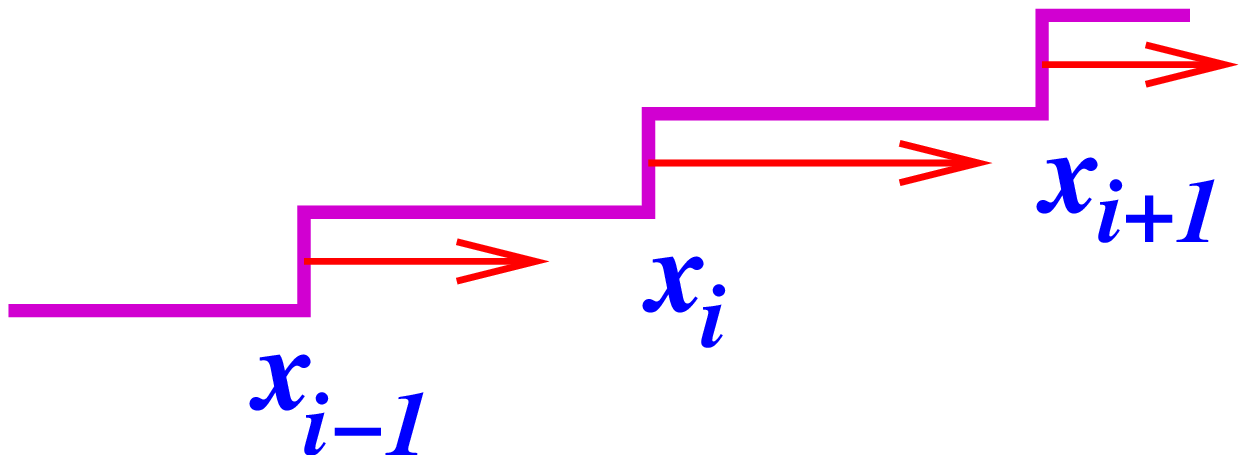}} 
\Caption{Sketch of a one-dimensional step train. Under sublimation,
ascending steps move to the right.}
\label{fig:steptrain}
\end{figure}

When step bunching is the dominant instability, the steps can
(to a first approximation) be assumed to be straight, and the 
problem reduces to the one-dimensional motion and interaction
of point-like steps. Figure \ref{fig:steptrain} illustrates
the situation for the case of \textit{sublimation}, where 
ascending steps move (on average) to the right. The equations
of motion for the steps can be obtained from the solution
of a one-dimensional moving boundary value problem for the
adatom concentration on the terraces. This procedure has been
reviewed in detail elsewhere \cite{Krug:Krug2005}. Here we start
the discussion directly from the nonlinear equations of motion, 
regarded as a (physically motivated) many-dimensional dynamical
system.

\subsection{Stability of step trains}

As a first orientation, suppose the velocity $\dot{x_i}$ of the
$i$'th step is the sum of contributions $f_+$ and $f_-$, which are
functions of the length of 
the leading terrace (in front of the step) and the trailing
terrace (behind the step), respectively, such that
\begin{equation}
\label{1Dstepmotion}
\frac{d x_i}{dt} = f_+(x_{i+1} - x_i) + f_-(x_i - x_{i-1})
\end{equation}
for the $N$ steps $i=1,...,N$, 
and periodic boundary conditions are employed.
Then a uniform step train of equally spaced steps,
\begin{equation}
\label{uniform}
x_i^{(0)} = i l + v t
\end{equation}
is always a solution, with $l$ denoting the step spacing
and $v = f_+(l) + f_-(l)$ the step speed. A straightforward
linear stability analysis of (\ref{1Dstepmotion}) reveals
that the solution (\ref{uniform}) is stable if 
\begin{equation}
\label{1Dstability}
\frac{d}{dx}[f_+(x) - f_-(x)] \vert_{x = l} > 0,
\end{equation}
and step bunching occurs when this condition is violated.

There are obviously different ways in which such an instability
can be realized. One possibility is that both contributions
on the right hand side of (\ref{1Dstepmotion}) are increasing
functions of the terrace size, but the contributon from the
trailing terrace is larger, i.e. the step motion is 
primarily driven from behind. This is the scenario first
described by Schwoebel and Shipsey \cite{Krug:Schwoebel1966,Krug:Schwoebel1969}, 
who pointed out that
the preferential attachment/detachment of adatoms from/to the
lower terrace bordering a step leads to step bunching during
sublimation. The mechanism for electromigration-induced step bunching
first described by Stoyanov \cite{Krug:Stoyanov1991} is of a similar nature.
We will return to this case in the following sections.

A different scenario was investigated by Kandel and Weeks
\cite{Krug:Kandel1992,Krug:Kandel1993}, who considered
a class of one-sided models with $f_- \equiv 0$ and a 
\textit{nonmonotonic} function $f_+$ of the form
\begin{equation}
\label{fplus}
f_+(x)=c x(x_0-x).
\end{equation}
This work was motivated by the physics of impurity-induced step bunching
during growth, where steps are slowed down by impurities that accumulate
on the terraces \cite{Krug:vanderEerden1989,Krug:Vollmer2008}. 
Larger terraces have been exposed to the impurity flux 
for longer times, which leads to a decrease of the step speed and ultimately
to its vanishing when $x = x_0$. The equidistant step train is stable for
$l < x_0/2$ and unstable for $l > x_0/2$. Perturbing a single step in an
unstable equidistant step train leads to a disturbance wave which
(because of the one-sided nature of the dynamics) travels backwards, leaving
behind a frozen configuration of step bunches separated by terraces of size
$x_0$. Varying the initial step spacing one finds a sequence of spatial bunching
patterns, which can be periodic, 
intermittent or chaotic\footnote{A similar scenario has been found in a 
model for sand ripple formation in an oscillatory flow 
\cite{Krug:Krug2001}.}.

\subsection{Strongly and weakly conserved step dynamics}

An important global characteristic of the step dynamics is 
the overall sublimation (or growth) rate of the crystal, which 
is given by
\begin{equation}
\label{R}
{\cal{R}} = \frac{1}{N} \sum_i \frac{dx_i}{dt}.
\end{equation}
We distinguish between \textit{strongly conserved} step dynamics
in which ${\cal{R}} = 0$, and \textit{weakly conserved} dynamics
where ${\cal{R}}$ is 
nonzero but independent of the step configuration\footnote{In 
\cite{Krug:Sato1999} only the \textit{strongly} conserved case is referred to as
``conserved''. The reason for our choice of nomenclature will become clear below
in Sect.\ref{Krug:Nonconserved}.}.
The latter case is realized during growth at relatively low temperature,
where desorption of adatoms can be neglected and therefore the
growth rate is completely determined by the external deposition 
flux \cite{Krug:Krug1997}. 

A generic model that incorporates the strongly and weakly conserved
situation is given by    
\begin{equation}
\label{consgen}
\frac{d x_i}{dt} = \gamma_+ \cdot (x_{i+1} - x_i) 
+ \gamma_- \cdot (x_i - x_{i-1}) 
+ U \cdot (2 f_i - f_{i+1} - f_{i-1})
\end{equation}
with
\begin{equation}
\label{forces}
f_i = \frac{l^3}{(x_i - x_{i-1})^3} - \frac{l^3}{(x_{i+1} - x_i)^3}.
\end{equation}
These equations were first written down by Liu and Weeks 
\cite{Krug:Liu1998} as a model
for electromigration-induced step 
bunching in the presence of sublimation\footnote{We will see below
in Sect.\ref{Krug:Nonconserved} that the weakly conserved form (\ref{consgen}) is in fact not 
really appropriate in the presence of sublimation.}.
In contrast to (\ref{1Dstepmotion}), here $\gamma_\pm$ and $U$
are constant coefficients multiplying the terms in parenthesis. 
Comparison with (\ref{1Dstepmotion}) shows that
$f_\pm$ are linear functions with slopes $\gamma_\pm$, such that
the stability condition reads $\gamma_+ > \gamma_-$. In addition
to the linear terms depending on the nearest neighbor step positions,
(\ref{consgen}) contains nonlinear next-nearest-neighbor contributions
arising from repulsive thermodynamic step-step interactions of entropic and
elastic origin \cite{Krug:Jeong99,Krug:Krug2005}, 
which drive the relaxation of the step train to its
(equidistant) equilibrium shape. 

The sublimation rate for the model (\ref{consgen}) is 
${\cal{R}} = (\gamma_+ + \gamma_-) l$, hence for strongly
conserved dynamics one has to set $\gamma_+ = - \gamma_-$.
This case is realized in electromigration-induced step bunching
without growth or sublimation \cite{Krug:Chang2006}. 
In the following we will focus on the weakly conserved
case, where ${\cal{R}} > 0$. It is then convenient to normalize
the time scale such that $\gamma_+ + \gamma_- = 1$, 
and to introduce the asymmetry parameter $b$ through
\cite{Krug:Popkov2005}
\begin{equation}
\label{b}
\gamma_+ = \frac{1 - b}{2}, \;\;\;
\gamma_- = \frac{1 + b}{2},
\end{equation}
such that step bunching occurs for $b > 0$. Together 
Eqs.(\ref{consgen},\ref{forces},\ref{b}) define 
a two-parameter family of nonlinear many-body problems which
have been investigated in detail in 
\cite{Krug:Popkov2005,Krug:Krug2005b,Krug:Popkov2006}. In the following
two sections some pertinent results of this study 
will be summarized.

\subsection{Continuum limit, traveling waves and scaling laws}

The analysis of the nonlinear dynamics of step bunches is greatly
simplified if it is possible to perform a continuum limit
of the problem, thus passing from the discrete dynamical
system (\ref{consgen}) to a partial differential equation
\cite{Krug:Krug2005,Krug:Krug1997a}.
Coarse graining the discrete equations of motion (\ref{consgen}),
one arrives first at a ``Lagrangian'' 
continuum description for the step positions $x_i$ or the 
terrace sizes $l_i = x_{i+1} - x_i$ by converting the 
layer index $i$ into a continuous surface height $h = i h_0$
(here $h_0$ denotes the height of an elementary step)
\cite{Krug:Chang2006,Krug:Fok2007}. In a second step this is 
transformed into an ``Eulerian'' evolution equation
for the surface height profile $h(x,t)$ or, equivalently, the
step density $m = \partial h/\partial x$ ,
which reads, for the model (\ref{consgen})
\cite{Krug:Popkov2005,Krug:Krug2005b}
\begin{equation}
\label{conteq}
\frac{\partial h}{\partial t} + \frac{\partial}{\partial x}
\left[ -\frac{b}{2m} - \frac{1}{6 m^3}
\frac{\partial m}{\partial x} + \frac{3 U}{2m} 
\frac{\partial^2 (m^2)}{\partial x^2} \right]
+ 1 = 0.
\end{equation}
To unburden the notation, we have normalized vertical and horizontal lengths
by setting $h_0 = l = 1$. In the weakly conserved case the evolution
law has the form of a continuity equation, with the corresponding
current given by the terms inside the square brackets.

\begin{figure}[t]
\def\capfrac{1}
\centerline{
\includegraphics[width=0.5\textwidth]{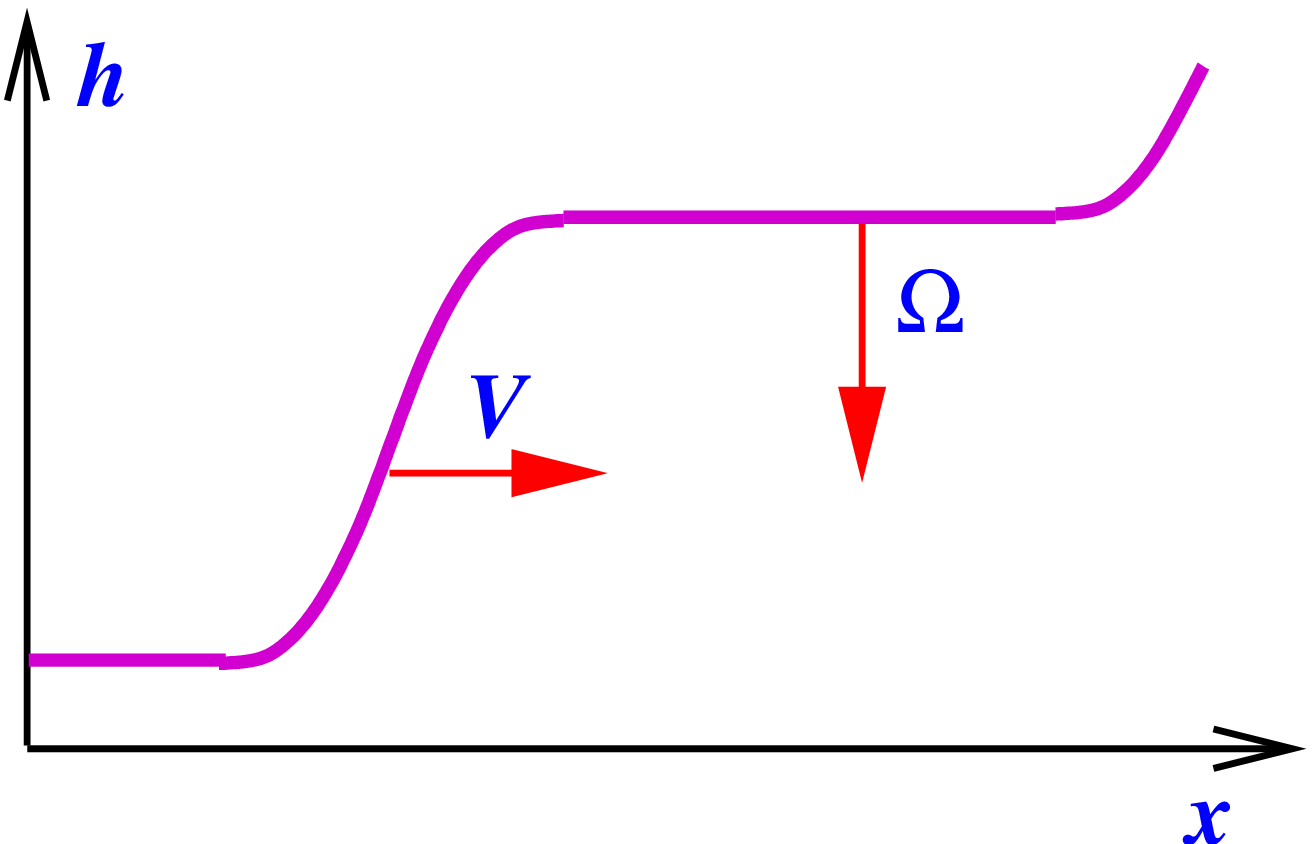}}
\Caption{Sketch of a moving step bunch.}
\label{fig:stepbunch}
\end{figure}

The solution $h(x,t) = x/l - t$ of (\ref{conteq}) is linearly unstable for
$b > 0$. The physically relevant nonlinear solutions take the form
of a generalized traveling wave,
\begin{equation}
\label{travel}
h(x,t) = f(x - Vt) - \Omega t
\end{equation}
as illustrated in Fig.\ref{fig:stepbunch}. The conserved nature of 
(\ref{conteq}) implies the sum rule
\begin{equation}
\label{sumrule}
\Omega + V = 1,
\end{equation}
but the individual values of the vertical and horizontal speed are
not fixed by the ansatz
\footnote{For the relation of this problem to the standard velocity selection problem for
traveling waves moving into unstable states see  
\cite{Krug:Slanina2005}.}. An analysis of periodic solutions of the 
discrete equations of motion shows that, under rather general conditions,
\begin{equation}
\label{V}
V \sim 1/N.
\end{equation}
Since the mean velocity of a single step is unity in the present units, this implies
that bunches move more slowly than steps. Similar to cars in a traffic jam, 
steps join the bunch from behind, move slowly through the bunch, and accelerate
into the \textit{outflow region} which separates one bunch from the 
next\footnote{Note however that traffic jams generally move in the direction
\textit{opposite} to the traffic flow \cite{Krug:Chowdhury2000,Krug:Helbing2001}.}.
\begin{figure}[t]
\def\capfrac{1}
\centerline{\includegraphics[width=0.6\textwidth]{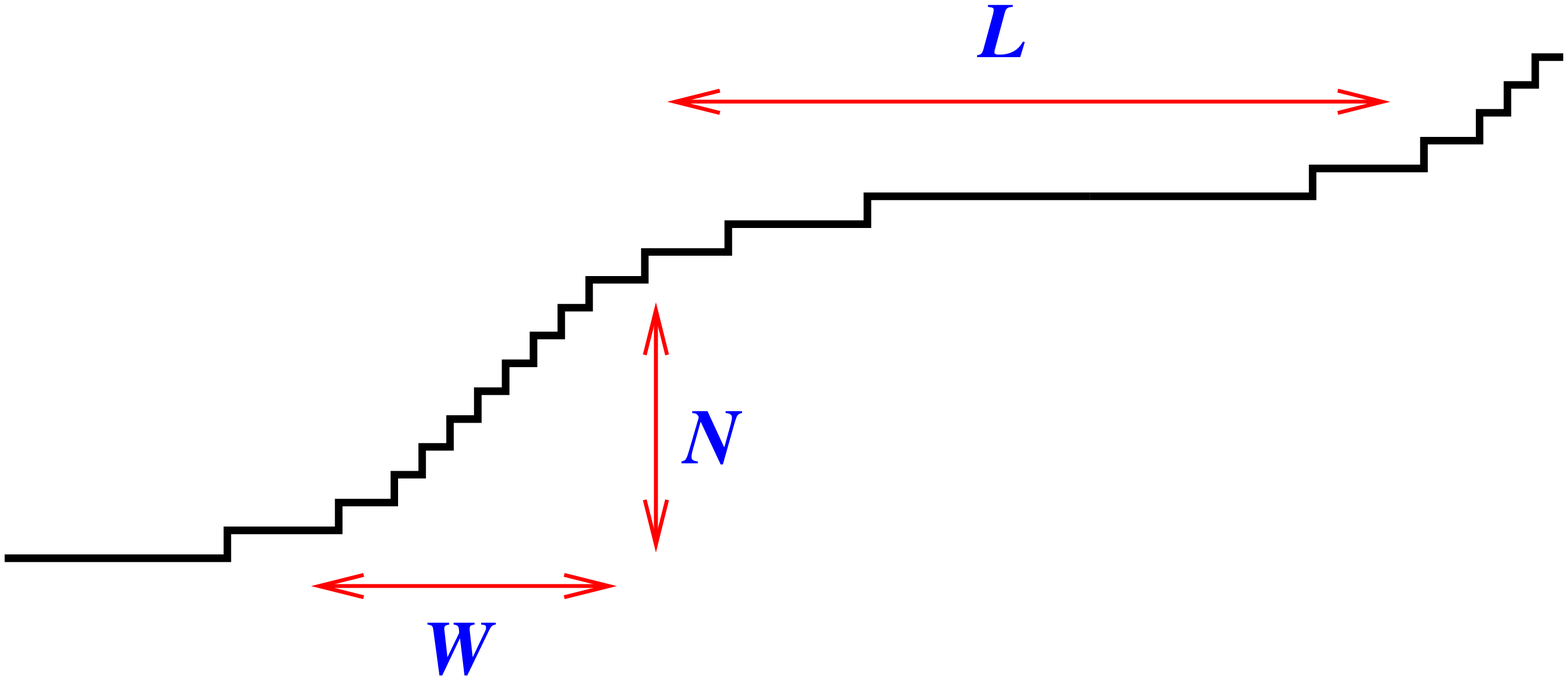}}
\Caption{Quantities characterizing the shape of a step bunch.}
\label{fig:stepbunch2}
\end{figure}

Inserting (\ref{travel}) into (\ref{conteq}) one arrives at a third order
nonlinear ODE, which can, to a large extent, be handled analytically
\cite{Krug:Popkov2005}. A key result are scaling laws 
\cite{Krug:Pimpinelli2002} for the shape of stationary bunches. As illustrated
in Fig.\ref{fig:stepbunch2}, the shape can be characterized by the bunch
width $W$ and the bunch spacing $L$, both of which are functions of the
number $N$ of steps in the bunch. The global constraint on the average
slope of the surface implies that $L \sim N$, but the bunch width
typically scales with a sublinear power of $N$, which implies that 
bunches become steeper as more steps are added. Related quantities
of interest are the minimal terrace size $l_\mathrm{min}$ in the bunch
and the size $l_1$ of the first terrace in the bunch. 
On the basis of the continuum equation (\ref{conteq}) one finds that,
asymptotically for large $N$
\cite{Krug:Popkov2005}
\begin{equation}
W \approx 4.1 (U N/b)^{1/3}, \;\;\;
l_\mathrm{min} \approx 2.4 (U/bN^2)^{1/3}, \;\;\;
l_1 \approx (2 U/bN)^{1/3},
\end{equation}
in good agreement with numerical simulations of the discrete model
\cite{Krug:Krug2005b}. Note that $W \sim N l_\mathrm{min}$, as one
would expect, but $l_1 \gg l_\mathrm{min}$. An
experimental study of the shapes of electromigration-induced 
step bunches on Si(111) is consistent with
$l_\mathrm{min} \sim N^{-2/3}$ \cite{Krug:Fujita1999}. 

\subsection{A dynamic phase transition}
\label{Krug:PhaseTransition}

As with any hydrodynamic description, the validity of the continuum
limit passing from (\ref{consgen}) to (\ref{conteq}) 
is restricted to step configurations
in which the step density is slowly varying on the scale of the
mean step spacing. To check the consistency of this assumption,
we consider the outflow region of the bunch, where the 
spacing between steps leaving the bunch becomes large and hence
the nonlinear interaction terms on the right hand side of (\ref{consgen})
can be neglected. We are thus left with the linear system
\begin{equation}
\label{dislin}
\frac{dx_i}{dt} = \frac{1-b}{2}(x_{i+1}-x_i) + \frac{1+b}{2}(x_i - 
x_{i-1}),
\end{equation}
which can be solved by the exponential traveling wave ansatz
\begin{equation}
\label{travelexp}
l_i \equiv x_{i+1} - x_i = A e^{Q(i+\Omega t)}.
\end{equation}
Inserting (\ref{travelexp}) into (\ref{dislin})
yields the relation
\begin{equation}
\label{bOmega}
b = \frac{\sinh Q - \Omega Q}{\cosh Q - 1} \approx 
\frac{\sinh Q - Q}{\cosh Q - 1}.
\end{equation}
where we have used that $\Omega \to 1$ for large bunches
according to (\ref{sumrule}) and (\ref{V}). 

\begin{figure}[t]
\def\capfrac{1}
\centerline{
\includegraphics[width=0.8\textwidth]{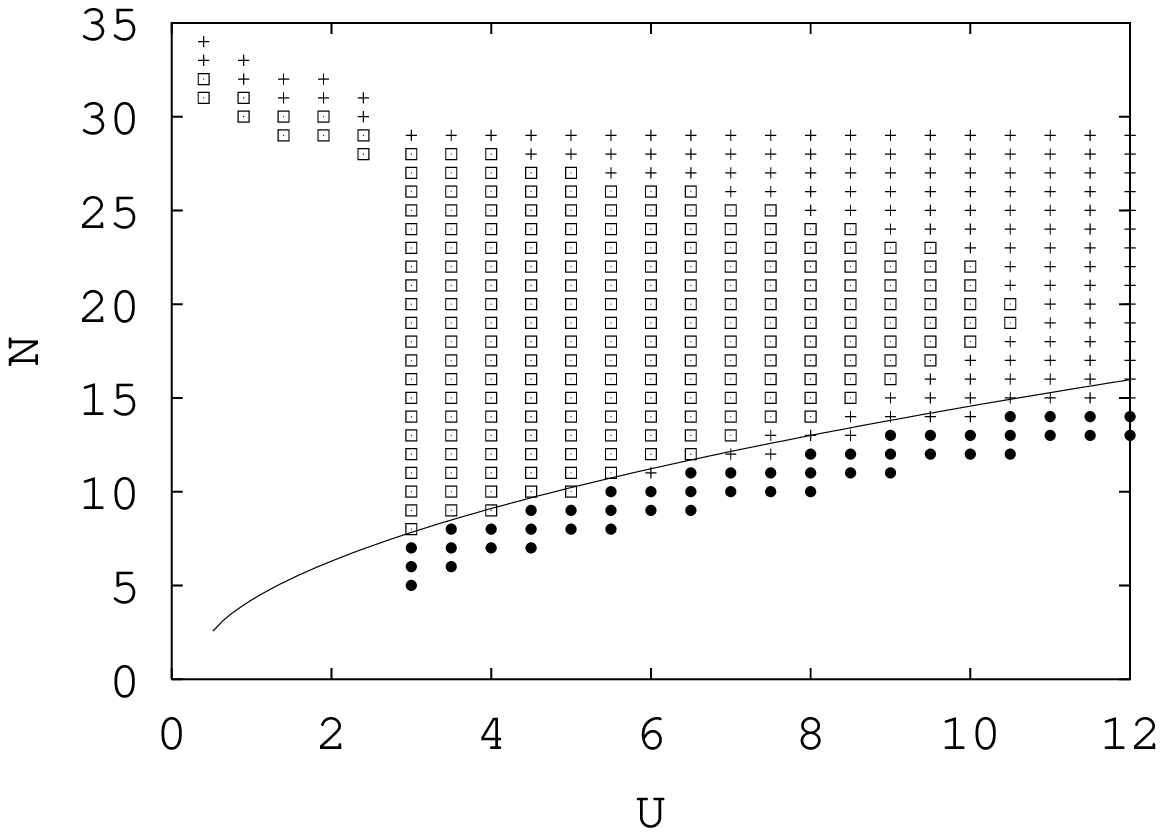}}
\Caption{Phase diagram for the behavior of step bunches
at $b = 11$. The line is the linear stability limit,
below which the equidistant step train is stable (full circles).
In the linearly unstable regime above this line bunches either
eject no steps (open squares) or they eject one step at a time
(crosses).}
\label{fig:popkov1}
\end{figure}

The step spacing is slowly varying when $Q \ll 1$, which according
to (\ref{bOmega}) requires $b \ll 1$. More strikingly, Eq.(\ref{bOmega})
has no solution when $b > 1$. At $b = 1$ the bunch undergoes a 
\textit{dynamic phase transition} which is reflected, among other things,
in the number of ``crossing'' steps between bunches: For $b < 1$
this number grows with $N$ as $\ln N$, whereas for $b > 1$ at most a single
step can reside between two bunches at one time \cite{Krug:Popkov2006}.

The physical origin of this change of behavior can be traced back
to the evolution equations (\ref{dislin}). For a step about to leave
the bunch the leading terrace is much larger than the trailing terrace,
$x_{i+1} -  x_i \gg x_i - x_{i-1}$, such that the right hand side of
(\ref{dislin}) is dominated by the first term, which is \textit{negative}
for $b > 1$. The linear term thus pushes the step back into the bunch,
and it can escape only thanks to 
the repulsive, nonlinear step-step interaction. Since the 
bunches become steeper with increasing size, the ability
of a bunch to eject crossing steps also depends on the number of steps
$N$ that it contains. 

\begin{figure}[t]
\def\capfrac{1}
\centerline{
\includegraphics[width=0.8\textwidth]{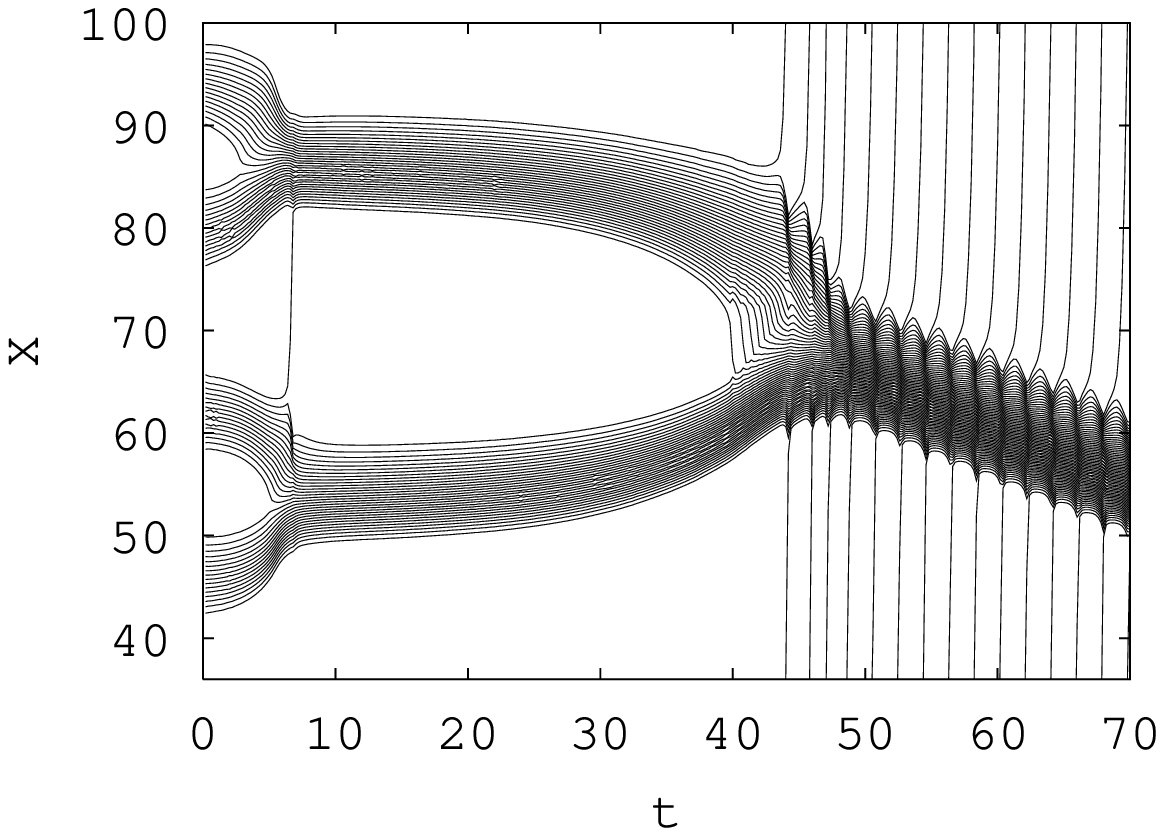}}
\Caption{Trajectories of 64 step evolving under the weakly
conserved dynamics (\ref{consgen}) with $b = 20$ and $U = 12$.
Step positions are shown in a frame moving with the mean step 
velocity. Initially the trajectories are horizontal, because
the entire bunch moves at the mean step speed.}
\label{fig:popkov2}
\end{figure}

The result of this interplay between linear
and nonlinear effects is the phase diagram in the 
$U$-$N$-plane depicted in Fig.\ref{fig:popkov1}.
At moderate values of $U$ it predicts a qualitative change in the 
behavior of bunches with increasing $N$. For small bunches the emission
of steps ceases completely, such that all steps constituting the bunch
move at the speed of the whole bunch and $V = 1$ in our units. Larger
bunches emit one step at a time. Figure \ref{fig:popkov2} shows the
transition between the two regimes in a time-dependent situation.
The initial condition consists of 4 small bunches of 16 steps each. 
These bunches initially merge in a hierarchical fashion without 
exchanging steps. This behavior is characteristic of \textit{strongly}
conserved step dynamics \cite{Krug:Sato1999,Krug:Chang2006},
which in our units corresponds to $b \to \infty$. After the last merger
the bunch enters the region in the phase diagram of Fig.\ref{fig:popkov1}
where step emission is possible, and correspondingly the overall bunch
motion slows down. It can also be seen that the emission of steps
is accompanied by a periodic ``breathing'' of the entire bunch
\cite{Krug:Popkov2006}. 

A rough estimate of experimental parameters
indicates that both regimes $b < 1$ and $b > 1$ can be accessed 
in experiments on electromigration-induced step bunching of the Si(111)
surface by varying the temperature \cite{Krug:Popkov2005}. 
The identification of the predicted
phase transition is however not straightforward because real steps 
can bend \cite{Krug:Thurmer1999}, 
thus invalidating the one-dimensional approximation used 
throughout this section.

\subsection{Coarsening}

The time evolution depicted in Fig.\ref{fig:popkov2} is an example
of \textit{coarsening}, a term that is generally used
to describe the (unlimited) increase of bunch size with time.
In many cases coarsening proceeds according to a power law,
\begin{equation}
\label{coarsening}
L \sim N \sim t^{n}
\end{equation}
defining the coarsening exponent $n$. Despite recent progress in the theory
of coarsening dynamics for one-dimensional fronts \cite{Krug:Politi2006},
a quantitative analysis of coarsening dynamics based on nonlinear continuum
equations such as (\ref{conteq}) seems still out of reach. Nevertheless,
heuristic arguments (to be explained below) in combination with numerical
\cite{Krug:Sato1999,Krug:Liu1998} and experimental 
\cite{Krug:Yang1996} evidence indicate that, as far
as the weakly conserved system (\ref{consgen}) including its 
strongly conserved limit is concerned, the coarsening
exponent is 
\begin{equation}
\label{coarsexp}
n = \frac{1}{2}
\end{equation}
under a wide range of conditions; in particular, the valuse of $n$
does not seem to be affected by the phase transition at $b = 1$
\cite{Krug:Ranguelov}. 

The first heuristic argument goes back to Chernov 
\cite{Krug:Chernov1961}, and it is based on the relation 
(\ref{V}) for the bunch velocity. The key assumption is that
$V$ is the only velocity scale in the problem, such that 
the velocity \textit{difference} between two bunches of similar
size $\sim N$ is also of order $\Delta V \sim 1/N$. The time required for two
bunches to merge is then of order $L/\Delta V \sim N^2$, and
(\ref{coarsexp}) follows. A weakness of this argument is that
it assumes coarsening to proceed by the merging of bunches, which
does not need to be true when bunches can exchange steps. 

The second argument, due to Liu and Weeks \cite{Krug:Liu1998},
is based on the generally conserved form of the continuum equation
for the height profile $h(x,t)$, which reads (in a frame where the
constant rate of sublimation has been subtracted)  
\begin{equation}
\label{conserved}
\frac{\partial h}{\partial t} + \frac{\partial j}{\partial x} = 0.
\end{equation}
Without further specifying the current $j$, Liu and 
Weeks assume the existence of a single lateral length scale
$\sim t^n$, such that both the height profile and the current
take on scaling forms
\begin{equation}
\label{scalings}
h(x,t) = t^n H(x/t^n), \;\;\; j(x,t) = J(x/t^n).
\end{equation}
Inserting (\ref{scalings}) into (\ref{conserved}) enforces (\ref{coarsexp}).
Similar scaling arguments have been advanced by Pimpinelli and coworkers
\cite{Krug:Pimpinelli2002}. 

Like the argument of Chernov, the ansatz (\ref{scalings}) is problematic
because the bunch spacing is \textit{not} the only length scale in
the system \cite{Krug:Krug2005a,Krug:Krug2005b}; 
for example, the bunch width $W$ defines a second 
(time-dependent) scale which
cannot obviously be ignored. An explicit counterexample where the existence
of an additional length scale leads to coarsening exponents which differ
from (\ref{coarsexp}) was presented in \cite{Krug:Slanina2005}.

\subsection{Nonconserved dynamics}
\label{Krug:Nonconserved}

\begin{figure}[t]
\def\capfrac{1}
\centerline{
\includegraphics[width=0.6\textwidth,angle=-90]{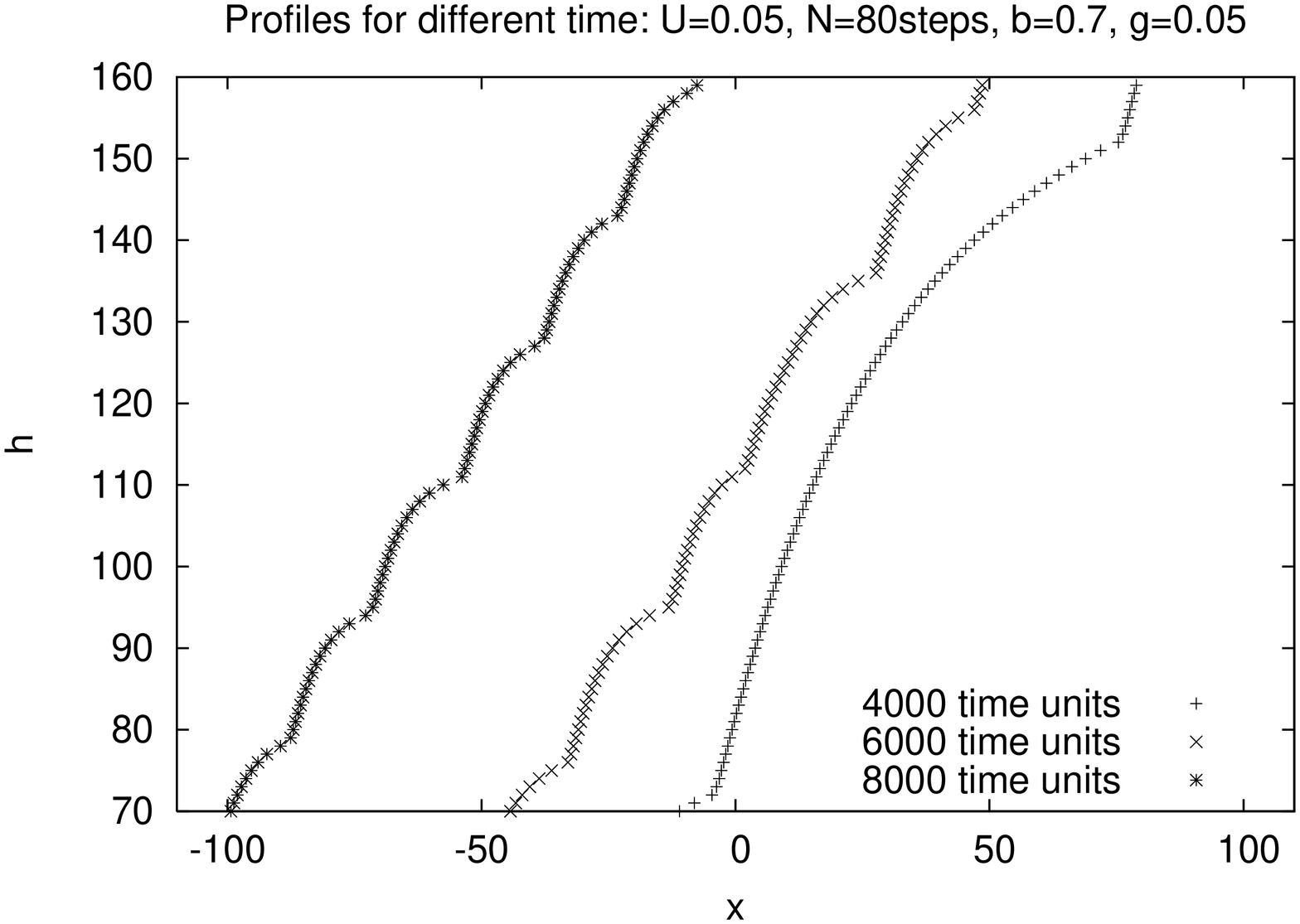}}
\Caption{Surface profiles generated with the nonconserved discrete model
(\ref{noncons}) with parameters $b=0.7$ and $U = g = 0.05$. The
initial condition is a single large bunch, which first relaxes into a 
quasi-stationary configuration and then breaks up into smaller bunches
after 4000 time steps. Height profiles at different times have
been shifted in the horizontal direction.}
\label{fig:Ivanov}
\end{figure}

In the presence of sublimation the rate of volume change 
(\ref{R}) couples to the step configuration, and therefore
the weakly conserved form of the discrete [Eq.(\ref{consgen})] and 
continuous [Eq.(\ref{conteq})] evolution equations is no longer
appropriate \cite{Krug:PierreLouis2003}. The minimal modification
of (\ref{consgen}) which takes account of this fact reads
\cite{Krug:Ivanov2007}
\begin{equation}
\label{noncons}
\frac{d x_i}{dt} = \left(1 + g f_i \right) \left[ \frac{1 + b}{2} (x_i - x_{i-1}) + \frac{1 - b}{2} (x_{i+1} - x_i) \right]
+ U (2 f_i - f_{i+1} - f_{i-1}),
\end{equation}
where the new dimensionless parameter $g$ is proportional to the
strength of the repulsive step-step interactions. On the linearized level
the introduction of the new term shifts the instability condition,
which now reads \cite{Krug:Fok2007,Krug:Ivanov2007} 
\begin{equation}
\label{instab}
b > 6 g.
\end{equation} 
The nonlinear consequences of the new term are quite dramatic. Numerical
simulations of (\ref{noncons}) \cite{Krug:Ivanov2008}, as well as of a more complicated
non-conserved model \cite{Krug:Sato1999} show that the coarsening of step 
bunches is \textit{arrested} when the bunches have reached a certain size. 
Correspondingly, a large initial step 
bunch evolving under the dynamics (\ref{noncons}) breaks up into smaller
bunches, as illustrated in Fig.\ref{fig:Ivanov}.  

The absence of (asymptotic) coarsening in the nonconserved case is consistent
with analyses in which \textit{weakly} nonlinear continuum equations (in the
sense of \cite{Krug:PierreLouis2005a}) are derived from the discrete step dynamics
close to the instability threshold, i.e. for $1 - 6 g/b \ll 1$ 
\cite{Krug:Sato1995,Krug:Misbah1996}. These equations typically display
spatio-temporal chaos or structure formation at a fixed length scale, but no
coarsening \cite{Krug:Krug2005a}. However, for strongly nonlinear continuum
equations similar to (\ref{conteq}), which are expected to apply when
$b \gg g$, such results are so far not available.

\subsection{Beyond the quasistatic approximation}

With few exceptions \cite{Krug:Ghez1988,Krug:Ghez1993,Krug:Keller1993}, most theoretical 
studies of step dynamics work in the \textit{quasistatic} approximation, which implies
that the dynamics of the diffusing adatoms on the terraces separating the steps is
assumed to be much faster than the step motion. As a consequence a step reacts instantaneously
to the motion of its neighbors, which mathematically leads to coupled first-order
equations for the step positions such as (\ref{consgen}). 

A simple and conceptually appealing
way of explicitly including the time scale of adatom dynamics was recently proposed
by Ranguelov and Stoyanov, who derived and studied a coupled system of two sets
of evolution equations, one for the terrace widths $l_i = x_{i+1} - x_i$ and one for
the (suitably parametrized) adatom concentration profile on the terraces.
Remarkably, in this setting the equidistant step train may undergo an instability
into a new dynamic phase characterized by step compression waves \cite{Krug:Ranguelov2007},
even if it would be completely stable in the quasistatic limit. The instability is caused
solely by the time delay that is introduced into the interaction between steps by the
finite time scale of the adatom dynamics, similar to the instabilities induced in 
follow-the-leader models of highway traffic
by the finite reaction time of drivers \cite{Krug:Chowdhury2000,Krug:Helbing2001}. 
In the presence of electromigration
and sublimation, the non-quasistatic model reproduces the main features of the phase
transition described above in Sect.\ref{Krug:PhaseTransition} \cite{Krug:Ranguelov2008}.

\section{Conclusions}

The fact that the evolution of nanostructures is intrinsically noisy 
is by now widely appreciated \cite{Krug:Williams2004}. 
In contrast, the role of deterministic nonlinear dynamics, in the
sense of dynamical systems theory, as a source of complex behavior 
is largely unexplored in this context. Here I have presented the results
of two case studies in which concepts from nonlinear dynamics
appear naturally in the analysis of the evolution
of surface nanostructures. In both cases surface steps constitute the
relevant degrees of freedom which, despite satisfying simple equations
of motion, can display a wide range of dynamic phenomena.
Many other systems, not discussed here, fit into the same framework;
an example of current interest is the thermal decay of nanoscale
mounds, either through
the periodic collapse of the top island \cite{Krug:Margetis2006}
or through the jerky rotation of a spiral step emanating from
a screw dislocation \cite{Krug:Ranganathan2005}.     
It should have become clear that much, perhaps most of the work in 
this field remains to be done.

\section*{Acknowledgements}

This chapter is based on joint work with 
Frank Hausser, Marian Ivanov, Philipp Kuhn, Vladislav Popkov, Marko Rusanen,
and Axel Voigt. I am grateful to Dionisios Margetis, Olivier Pierre-Louis,
Alberto Pimpinelli, Paolo Politi, Bogdan Ranguelov,
Stoyan Stoyanov, Vesselin Tonchev and John D. Weeks for useful interactions, and to
DFG for support within SFB 616 \textit{Energy dissipation at surfaces}
and project KR 1123/1-2.

\newpage

\setlength{\bibindent}{4mm} 
